\documentclass[9pt,conference]{IEEEtran}
\usepackage{dcase2025}

\usepackage{bm} 

\usepackage{dcase2025,amsmath,graphicx,url,times,booktabs, tabularx}

\usepackage{float}

\usepackage{wrapfig}
\usepackage{lipsum}
\usepackage{threeparttable}

\title{An Entropy-Guided Curriculum Learning Strategy for Data-Efficient Acoustic Scene Classification under Domain Shift}

\name{Peihong Zhang, Yuxuan Liu*\thanks{Peihong Zhang and Yuxuan Liu contributed equally to this work.}, Zhixin Li, Rui Sang, Yiqiang Cai, Yizhou Tan, Shengchen Li}

\address{School of Advanced Technology, Xi’an Jiaotong-Liverpool University, Suzhou, China\\
\{Peihong.Zhang20, Yuxuan.Liu2204, Zhixin.Li22, Rui.Sang22, \\ Yiqiang.Cai21, Yizhou.Tan22\}@student.xjtlu.edu.cn\\
Shengchen.Li@xjtlu.edu.cn}

\begin{document}

\maketitle

\begin{abstract}
Acoustic Scene Classification (ASC) faces challenges in generalizing across recording devices, particularly when labeled data is limited. The DCASE 2024 Challenge Task 1 highlights this issue by requiring models to learn from small labeled subsets recorded on a few devices. These models need to then generalize to recordings from previously unseen devices under strict complexity constraints. While techniques such as data augmentation and the use of pre-trained models are well-established for improving model generalization, optimizing the training strategy represents a complementary yet less-explored path that introduces no additional architectural complexity or inference overhead. Among various training strategies, curriculum learning offers a promising paradigm by structuring the learning process from easier to harder examples. In this work, we propose an entropy-guided curriculum learning strategy to address the domain shift problem in data-efficient ASC. Specifically, we quantify the uncertainty of device domain predictions for each training sample by computing the Shannon entropy of the device posterior probabilities estimated by an auxiliary domain classifier. Using entropy as a proxy for domain invariance, the curriculum begins with high-entropy samples and gradually incorporates low-entropy, domain-specific ones to facilitate the learning of generalizable representations. Experimental results on multiple DCASE 2024 ASC baselines demonstrate that our strategy effectively mitigates domain shift, particularly under limited labeled data conditions. Our strategy is architecture-agnostic and introduces no additional inference cost, making it easily integrable into existing ASC baselines and offering a practical solution to domain shift.
\end{abstract}


\section{Introduction}
\label{sec:intro}


Acoustic Scene Classification (ASC) aims to recognize the environment from an audio segment and serves as a fundamental task in computational auditory analysis \cite{barchiesi2015acoustic,cai2024tf}. A major challenge in real-world ASC systems is domain shift caused by variations in recording devices, which can significantly degrade model generalization to unseen devices. The challenge is further exacerbated when only limited labeled training data is available. The DCASE 2024 Challenge Task 1 \cite{schmid2024data} explicitly targets this scenario by requiring systems to be trained on small fractions (as low as 5\%) of labeled audio from a limited set of devices and evaluated on recordings from previously unseen devices. In addition, participating models must adhere to strict complexity constraints, including a maximum of 128 kB of parameters and 30 MMACs per second of audio.

\begin{figure}[t]
    \centering
    \includegraphics[width=0.33\textwidth]{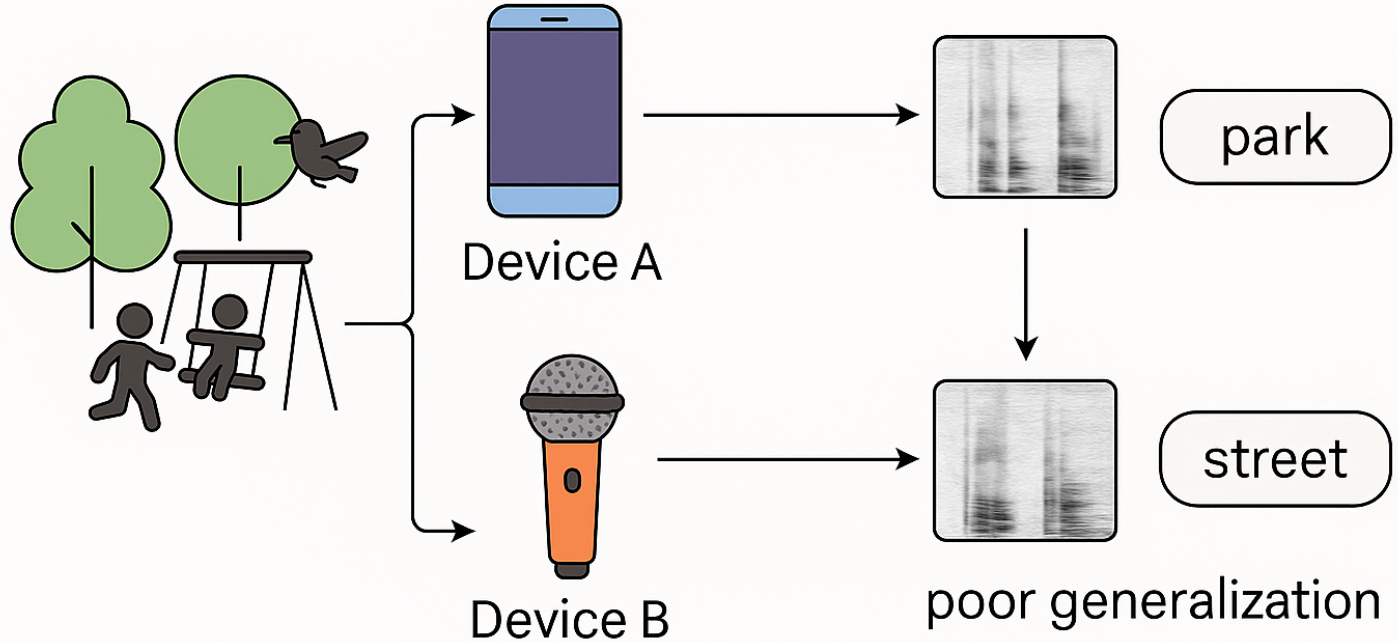} 
    \caption{Example of domain shift in ASC caused by device variability. Although the two spectrograms represent the same scene, device-specific characteristics cause noticeable differences. As a result, a model trained on one device may fail to correctly classify inputs from another, illustrating poor generalization.}
    \label{fig:domain_shift}
\end{figure}

Domain shift becomes particularly challenging when labeled training data is limited, as models struggle to learn domain-invariant representations from limited supervision \cite{tan2025domain}. Figure~\ref{fig:domain_shift} illustrates this phenomenon in ASC, where spectrograms of the same scene recorded by different devices exhibit noticeable discrepancies. 

To address such challenges, existing approaches often turn to data augmentation strategies—such as Freq-MixStyle~\cite{schmid2022cp} and device impulse response simulation~\cite{cai2023dcase2023}—or leverage pretrained models~\cite{cai2024dcase2024} to enhance model generalization across recording devices. These techniques improve feature diversity and representation quality but may also introduce additional complexity \cite{perez2017effectiveness,tan2019efficientnet}, such as increased model size, reliance on external resources, or inference overhead. 

While these approaches are effective, they focus on modifying the data or the initial state of the model. The training curriculum itself—the very order in which data is presented to the model—remains a relatively underexplored dimension for enhancing domain generalization \cite{wang2022generalizing}. This observation motivates our central question: Can we devise a lightweight, model-agnostic training strategy that optimizes the learning trajectory itself to build domain-invariant representations with limited supervision?




To address this challenge, we draw inspiration from Curriculum Learning \cite{bengio2009curriculum}, a training paradigm that organizes the presentation of training data from easy to hard, thereby optimizing the learning trajectory. This approach is particularly well-suited for data-limited domain shift scenarios, as it facilitates the progressive acquisition of generalizable representations \cite{liu2022less}. By introducing less confounded, domain-invariant samples early in training, the model can establish a robust feature foundation before gradually incorporating harder, domain-specific examples. Such staged learning has been shown to enhance generalization in a variety of tasks \cite{zhang2017curriculum, wang2024curriculum}. Building on this idea, we propose an entropy-guided curriculum learning strategy that complements existing approaches without modifying the ASC model architecture. Specifically, we employ an auxiliary domain classifier to estimate the device posterior probabilities for each training sample. The Shannon entropy of this distribution serves as a proxy for domain invariance, based on the intuition that samples with high entropy—those that confuse the device classifier—are more likely to be device-agnostic and thus easier for learning generalizable features \cite{arsenos2024uncertainty}. The curriculum therefore begins with these domain-invariant samples to establish a robust feature foundation, and gradually introduces harder, domain-specific examples to refine decision boundaries.




To assess the effectiveness of our approach, we conduct evaluations across multiple ASC systems submitted to the DCASE 2024 Challenge Task 1, with diverse model architectures and training setups. Experimental results show that our method consistently improves cross-device performance under low-resource conditions. These findings demonstrate the potential of entropy-guided curriculum learning as a lightweight, architecture-agnostic, and easily integrable strategy for improving domain generalization in ASC.

\begin{figure*}[t]
  \centering
  \includegraphics[width=0.73\textwidth]{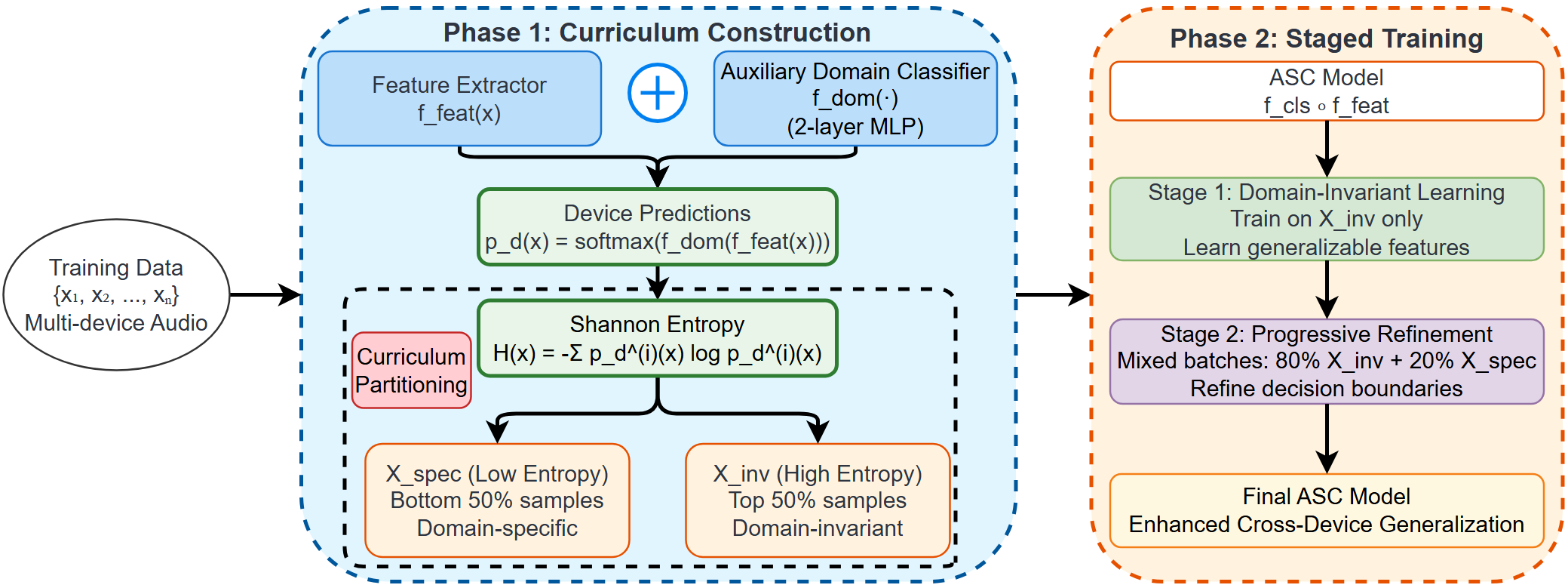}
  \caption{Overview of the proposed entropy-guided curriculum learning framework. In Phase~1, an auxiliary domain classifier estimates the device posterior distribution for each training sample, from which entropy is computed to quantify domain uncertainty. Based on entropy values, samples are partitioned into domain-invariant ($X_{\mathrm{inv}}$) and domain-specific ($X_{\mathrm{spec}}$) subsets. In Phase~2, the ASC model is initially trained on $X_{\mathrm{inv}}$ to promote domain-invariant feature learning, followed by a staged refinement phase using mixed batches composed of both $X_{\mathrm{inv}}$ and $X_{\mathrm{spec}}$, thereby enhancing robustness to domain shift.}

  \label{fig:entropy-curriculum-flowchart}
\end{figure*}

\section{Background}

\subsection{Domain Shift in Data-Efficient ASC}

Domain shift, particularly induced by device variability, is a well-established challenge in Acoustic Scene Classification (ASC)~\cite{tan2024acoustic}. Differences in microphone frequency response and spatial positioning introduce systematic shifts in audio characteristics across devices~\cite{bai2024description}. These discrepancies can significantly impair cross-device generalization, as models trained on one device distribution often fail to transfer effectively to others \cite{morocutti2023device}. The impact is especially pronounced under limited labeled data, where insufficient device diversity prevents the model from learning robust, domain-invariant representations~\cite{schmid2024data}.

When training data covers a diverse range of scenes and recording devices, domain-invariant features can emerge naturally, as the model learns to disentangle scene-relevant cues from device-specific artifacts~\cite{schmid2022cp}. In contrast, under data-efficient settings, such coverage is lacking, making the learning process prone to overfitting device characteristics.  For instance, if recordings corresponding to a specific scene are predominantly collected using a single device, the model may implicitly associate the spectral properties of that device with the scene label \cite{yan2024semi}. Such spurious correlations introduce domain leakage and compromise robustness \cite{tan2024acoustic,liu2025maiainpaintingbasedapproachmusic}, as the model relies on incidental device-dependent patterns rather than genuine scene-discriminative features. Furthermore, early exposure to domain-specific samples may bias the internal representation space of the model, steering the learning process toward device-dependent features~\cite{hacohen2019power}.

The above findings suggest that, under limited supervision, it may be beneficial to control the order in which training samples are presented. By prioritizing domain-invariant examples early in training, the model may better avoid overfitting to device-specific cues. This intuition motivates the use of a curriculum-based learning schedule to improve generalization without altering model architecture.

\subsection{Entropy-Guided Curriculum Learning}

Curriculum Learning (CL) is a training paradigm inspired by the human learning process, in which knowledge is acquired progressively—from simple concepts to more complex ones \cite{bengio2009curriculum}. By exposing the model to easier examples in the early stages of training and gradually increasing sample difficulty, CL has been shown to improve optimization stability and generalization, particularly in settings with limited labeled data \cite{hacohen2019power,jiang2018mentornet}.

A key component of curriculum learning is the definition of sample difficulty, which determines the order in which training examples are presented. Existing curricula commonly rely on heuristics such as loss magnitude~\cite{jiang2018mentornet}, prediction confidence~\cite{wang2019symmetric}, classification margin~\cite{hacohen2019power}, or uncertainty estimates derived from Bayesian inference~\cite{lakshminarayanan2017simple}. While these strategies are effective in many supervised learning tasks, their core assumption—that task performance is a valid proxy for sample difficulty—becomes untenable in the context of domain shift. The reason is that task performance is an inherently ambiguous metric: a sample with low classification loss may be either genuinely simple and domain-invariant, or deceptively easy due to spurious, domain-specific cues. This ambiguity renders such a metric an unsound foundation for any curriculum whose objective is to improve generalization.

To address this limitation, we propose to define sample difficulty in terms of domain uncertainty rather than task performance. The core idea is to prioritize training on examples that are minimally influenced by domain-specific features, and are therefore more likely to generalize across domains. To quantify domain uncertainty, we employ an auxiliary domain classifier to estimate the device posterior distribution for each sample. The Shannon entropy of this distribution serves as a proxy for domain invariance: samples with higher entropy indicate greater ambiguity in device identity, suggesting a lower reliance on device-dependent features \cite{arsenos2024uncertainty}. By organizing data according to this entropy-based measure, we construct a curriculum that gradually transitions from domain-invariant to domain-specific examples, thereby facilitating the learning of more generalizable representations. While curriculum learning has been explored in domain generalization for tasks like semantic segmentation \cite{zhang2017curriculum} and fault diagnosis \cite{wang2024curriculum}, its application to acoustic scene classification remains largely unexplored.

\section{Proposed Method: Entropy-Guided Curriculum Learning}

\subsection{Overview and Problem Setup}

We propose an entropy-guided curriculum learning strategy to enhance the generalization capability of ASC models under domain shift. The strategy comprises three stages: (1) quantifying domain uncertainty using an auxiliary device classifier, (2) constructing a training curriculum by ranking samples based on entropy, (3) performing staged model training from domain-invariant to domain-specific samples.

Each ASC model consists of a feature extractor \( f_{\text{feat}} \) and a scene classifier \( f_{\text{cls}} \). To construct the curriculum, we substitute \( f_{\text{cls}} \) with a lightweight device classifier \( f_{\text{dom}} \), which is trained to predict device identities. For each training sample \( x \), the device classifier produces a posterior probability distribution over device labels. We compute the Shannon entropy \( H(x) \) of this distribution to measure domain ambiguity. A higher entropy indicates greater uncertainty in device identity, implying stronger domain invariance.

We rank all samples by entropy and split them at the median: the top 50\% form the domain-invariant subset \( \mathcal{X}_{\text{inv}} \), while the remaining form the domain-specific subset \( \mathcal{X}_{\text{spec}} \). Training begins exclusively on \( \mathcal{X}_{\text{inv}} \) and progressively incorporates \( \mathcal{X}_{\text{spec}} \) through mixed mini-batches. This method requires no changes to model architecture or hyperparameters, ensuring seamless integration into existing ASC models.



\subsection{Entropy-Based Curriculum Construction}

To estimate domain uncertainty, we first freeze the feature extractor \( f_{\text{feat}} \) and train an auxiliary device classifier \( f_{\text{dom}} \) based on it. Given an sample \( x \), the classifier produces a posterior distribution over device:
\begin{equation}
\mathbf{p}_d(x) = \text{softmax}(f_{\text{dom}}(f_{\text{feat}}(x)))
\end{equation}

The classifier is trained using device labels and the standard cross-entropy loss. After training, we compute the Shannon entropy of the predicted distribution for each sample:
\begin{equation}
H(x) = -\sum_{i=1}^{D} p_d^{(i)}(x) \log p_d^{(i)}(x)
\end{equation}
where \( D \) is the number of recording devices. Higher entropy indicates greater ambiguity in device identity, which implies that the sample is less influenced by device-specific characteristics and therefore exhibits stronger domain invariance.





Let \( \{x_i\}_{i=1}^N \) denote the training set. We sort all samples based on their entropy values and partition them into two curriculum subsets:
\begin{align}
\mathcal{X}_{\text{inv}} &= \left\{x_i \in \mathcal{D}_{\text{train}} \,\middle|\, \text{rank}(H(x_i)) \leq \lfloor 0.5N \rfloor \right\} \\
\mathcal{X}_{\text{spec}} &= \mathcal{D}_{\text{train}} \setminus \mathcal{X}_{\text{inv}}
\end{align}
We adopt a median-based entropy split as a simple yet effective proxy for curriculum ordering, selecting the top 50\% high-entropy samples as domain-invariant. This empirical threshold ensures a balanced curriculum distribution between domain-invariant and domain-specific examples. While this strategy is effective, we acknowledge that sample difficulty may vary continuously. Future work will explore adaptive or weighted curriculum scheduling schemes.

In the second training stage, training mini-batches \( \mathcal{B} \subset \mathcal{D}_{\text{train}} \) are constructed using a fixed-ratio sampling strategy:
\begin{equation}
\mathcal{B} = \mathcal{B}_{\text{inv}} \cup \mathcal{B}_{\text{spec}}, \quad
|\mathcal{B}_{\text{inv}}| = \lfloor 0.8B \rfloor, 
\quad |\mathcal{B}_{\text{spec}}| = B - |\mathcal{B}_{\text{inv}}|
\end{equation}

The gradual incorporation of domain-specific samples facilitates stable convergence and improves generalization across unseen devices.

\subsection{Curriculum-Guided Model Training}

After curriculum construction, we reinstate the original scene classifier \( f_{\text{cls}} \) and train the full model \( f_{\text{cls}} \circ f_{\text{feat}} \).

Training is performed using the standard cross-entropy loss:
\begin{equation}
\mathcal{L}_{\text{ASC}} = -\sum_{i=1}^{C} y^{(i)} \log \hat{y}^{(i)}, \quad \hat{y} = \text{softmax}(f_{\text{cls}}(f_{\text{feat}}(x)))
\end{equation}
where \( C \) is the number of scene classes and \( y \) is the one-hot label.

The model is trained in two stages:
\begin{itemize}
    \item \textbf{Stage 1:} The model is first trained exclusively on \( \mathcal{X}_{\text{inv}} \) to facilitate the learning of domain-invariant representations.
    \item \textbf{Stage 2:} To enhance cross-device generalization, the model is progressively exposed to domain-specific inputs by constructing each mini-batch with a fixed 80:20 ratio of samples from \( \mathcal{X}_{\text{inv}} \) and \( \mathcal{X}_{\text{spec}} \), respectively. This controlled inclusion enables the model to adapt to domain-specific cues while preserving the domain-invariant features learned during Stage 1.

\end{itemize}

The transition from Stage 1 to Stage 2 is triggered when the validation loss on \( \mathcal{X}_{\text{inv}} \) stops improving beyond a predefined threshold for several consecutive epochs. This reflects the core principle of curriculum learning: once the model has sufficiently learned from easier, domain-invariant samples, it is gradually exposed to more challenging, domain-specific data. Throughout both stages, we strictly follow the original training configurations of each ASC baseline, including the optimizer, learning rate schedule, and batch size. This ensures fair comparison and highlights the plug-and-play compatibility of our method. By progressively shifting the training focus, the staged curriculum promotes robust representation learning and improves generalization to unseen devices.








\section{Experiment}





\subsection{Experimental Setup}

\subsubsection{Datasets}
Our experimental evaluation is performed using the DCASE 2024 Task 1 dataset \cite{Heittola2020}, a benchmark derived from the TAU Urban Acoustic Scenes 2022 Mobile development set for the Acoustic Scene Classification (ASC) task. The dataset consists of one-second audio clips representing ten distinct acoustic scenes, recorded across 12 European cities with multiple real and simulated mobile devices. In total, the development set contains 230,350 audio segments, amounting to approximately 64 hours of audio.

As detailed in Table~\ref{tab:dataset_distribution}, the dataset is partitioned into training and testing sets with a specific device distribution. The training set includes labeled data from real devices (A, B, C) and simulated devices (S1, S2, S3). A significant characteristic of the training data is its imbalance, with Device A contributing a substantial majority of the segments. The test partition comprises data from all training devices and, critically, introduces three previously unseen simulated devices (S4, S5, S6) to assess the generalization capabilities of the models.

\begin{table}[H]
\centering
\caption{Distribution of DCASE 2024 Task 1 Development Dataset}
\label{tab:dataset_distribution}
\begin{tabular}{p{1.3cm} p{1cm} p{1.45cm} p{1.45cm} p{1.45cm}}
\toprule
\textbf{Devices} & \textbf{Type} & \textbf{Total segments} & \textbf{Train segments} & \textbf{Test segments} \\
\midrule
A & Real & 144,000 & 102,150 & 3,300 \\
B & Real & 10,780 & 7,490 & 3,290 \\
C & Real & 10,770 & 7,480 & 3,290 \\
S1, S2, S3 & Simulated & 3 $\times$ 10,800 & 3 $\times$ 7,500 & 3 $\times$ 3,300 \\
S4, S5, S6 & Simulated & 3 $\times$ 10,800 & - & 3 $\times$ 3,300 \\
\midrule
\textbf{Total} & & \textbf{230,350} & \textbf{139,620} & \textbf{29,680} \\
\bottomrule
\end{tabular}

\end{table}








\begin{table*}[t]
\centering
\caption{Classification accuracy (\%) on the DCASE2024 Task 1 evaluation set across different training data splits. Each system is compared with and without our proposed curriculum learning strategy. The last two columns report accuracy on seen and unseen devices under the 5\% training condition.}
\label{tab:main_results_seen_unseen_5percent}
\begin{tabular}{lccccc|cc}
\toprule
\textbf{System} & \textbf{5\%} & \textbf{10\%} & \textbf{25\%} & \textbf{50\%} & \textbf{100\%} & \textbf{Seen (5\%)} & \textbf{Unseen (5\%)} \\
\midrule

\textbf{DCASE2024 Baseline} \cite{Schmid2023} & 44.00 & 46.95 & 51.47 & 54.40 & 56.84 & 45.3 & 42.4 \\
+ our strategy & \textbf{46.30} & \textbf{49.10} & \textbf{52.80} & \textbf{55.20} & \textbf{57.00} & \textbf{47.5} & \textbf{44.0} \\

\midrule

\textbf{Cai\_XJTLU} \cite{cai2024dcase2024} & 48.91 & 53.16 & 58.09 & 59.47 & 62.12 & 50.6 & 46.7 \\
+ our strategy & \textbf{51.50} & \textbf{55.10} & \textbf{59.50} & \textbf{61.25} & \textbf{63.20} & \textbf{52.3} & \textbf{49.3} \\

\midrule

\textbf{Han\_SJTUTHU} \cite{Bing2024} & 54.35 & 56.69 & 59.09 & 60.38 & 61.82 & 55.6 & 52.7 \\
+ our strategy & \textbf{56.60} & \textbf{58.25} & \textbf{60.40} & \textbf{61.65} & \textbf{62.20} & \textbf{57.3} & \textbf{55.2} \\

\bottomrule
\end{tabular}
\end{table*}

To investigate performance under data-limited conditions, we strictly adhere to the official low-resource protocol of the DCASE 2024 Challenge. This protocol provides five predefined training subsets drawn from the multi-device data (A, B, C, S1, S2, S3), which contain 5\%, 10\%, 25\%, 50\%, and 100\% of the total labeled data, respectively. All models in our study are trained exclusively on these specified low-resource subsets.

\subsubsection{Evaluation Metrics}
Model performance is primarily evaluated using the official DCASE 2024 Task 1 metric: class-wise average accuracy. This metric computes the mean of per-class accuracies, ensuring a balanced performance assessment across all scene categories. It is defined as:
\begin{equation}
\text{Accuracy}_{\text{avg}} = \frac{1}{C} \sum_{c=1}^{C} \frac{N_c^{\text{correct}}}{N_c} \label{eq:accuracy}
\end{equation}
where \( C \) denotes the total number of scene classes, \( N_c \) represents the total number of samples belonging to class \( c \), and \( N_c^{\text{correct}} \) is the count of correctly classified samples for that specific class.

\subsubsection{Baseline Systems}

To validate the general applicability of our entropy-guided curriculum strategy, we evaluate three representative baseline systems from DCASE2024:

\begin{enumerate}
  \item \textbf{DCASE2024 Official Baseline}~\cite{Schmid2023}: employs CP-Mobile, a CNN with residual inverted bottleneck blocks. It utilizes knowledge distillation from a teacher ensemble (PaSST and CP-ResNet), with augmentations like Freq-MixStyle and device impulse response simulation.
  
  \item \textbf{Cai\_XJTLU}~\cite{cai2024dcase2024} (ranked 4th): adopts a teacher–student approach with a pretrained PaSST teacher and a TF-SepNet-64 student trained via knowledge distillation. Training includes SpecAugment, Mixup, noise addition, and pseudo-labeling.
  
  \item \textbf{Han\_SJTUTHU}~\cite{Bing2024} (ranked 1st): uses a student–teacher framework with SSCP-Mobile (a spatially separable CP-Mobile variant) as the student, and a PaSST teacher ensemble. Techniques include knowledge distillation, SpecAugment, Freq-MixStyle, and model pruning.
\end{enumerate}




\subsubsection{Implementation Details}
Our entropy-guided curriculum learning strategy is implemented in an architecture-agnostic manner. Accordingly, we follow each baseline system's original training protocol, including optimizer, learning rate schedule, batch size, and other hyperparameters.

\begin{itemize}
    \item \textbf{Auxiliary Domain Classifier:} A lightweight two-layer Multi-Layer Perceptron (MLP) is used to predict device identities from features extracted by the frozen primary encoder \( f_{\text{feat}} \). The MLP comprises one hidden layer with ReLU activation and a softmax output layer, and is optimized using Adam with a learning rate of \( 1 \times 10^{-4} \).

    \item \textbf{Curriculum-Guided Training:} The ASC model is trained in two stages as described in Section~3.3. The transition to Stage 2 is based on validation performance on \( \mathcal{X}_{\text{inv}} \), and training follows the fixed 80:20 sampling strategy. All models are trained under their respective official settings to ensure fair comparison.
\end{itemize}

\subsection{Results and Analysis}

Table~\ref{tab:main_results_seen_unseen_5percent} shows the classification accuracy on the DCASE2024 Task 1 evaluation set across varying proportions of labeled training data. Performance improvements introduced by our strategy are particularly notable under data-limited conditions (e.g., 5\%–25\% of training data), where models are more susceptible to domain shift.

To better assess the impact of our method on domain generalization, we decompose performance under the 5\% training condition into seen-device and unseen-device subsets, where domain shift is most pronounced. As shown in the additional columns of Table~\ref{tab:main_results_seen_unseen_5percent}, our strategy consistently improves both subsets, with notably larger gains on unseen devices. For instance, in the Cai\_XJTLU, accuracy on unseen devices increases from 46.7\% to 49.3\% (+2.6\%), whereas the improvement on seen devices is comparatively modest (+1.7\%).



\begin{figure}[t]
    \centering
    \includegraphics[width=0.98\linewidth]{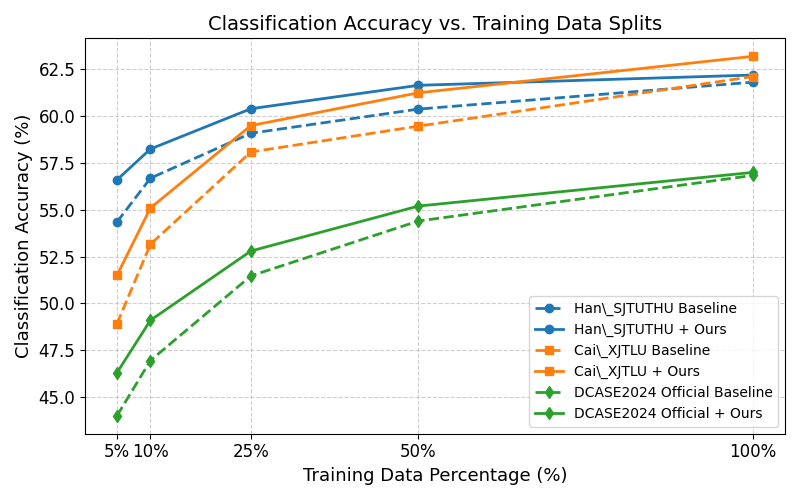}
    \caption{Classification accuracy across varying training data splits for different baseline systems, with and without our proposed strategy.}
    \label{fig:result_curve}
\end{figure}

We further illustrate these improvements in Fig.~\ref{fig:result_curve}, which presents accuracy curves across varying data splits. The curves clearly show that our curriculum strategy consistently enhances model generalization across different data regimes. Specifically, improvements gradually diminish as more training data becomes available, which is consistent with the expectation that abundant labeled data already facilitates effective learning of domain-invariant features, thereby reducing the relative benefits of training strategies.

Overall, these results confirm the efficacy of our entropy-guided curriculum learning strategy in improving ASC model generalization under domain shift, particularly in scenarios characterized by limited labeled training data. Importantly, the proposed method does not introduce additional inference overhead or architectural complexity, making it a practical solution for ASC tasks under domain shift.

\section{Conclusion}

We proposed an entropy-guided curriculum learning strategy to address domain shift in ASC under limited labeled data. By employing an auxiliary domain classifier to estimate domain uncertainty via entropy, the training process is structured to progress from domain-invariant to domain-specific samples, facilitating the acquisition of robust and generalizable representations. Experimental results on multiple DCASE2024 Task 1 baseline systems demonstrate consistent performance gains across training regimes, with particularly notable improvements in low-resource scenarios. The proposed strategy is architecture-agnostic, incurs no additional inference overhead, and integrates seamlessly with existing ASC models.




\clearpage
\bibliographystyle{IEEEtran}
\bibliography{refs}







\end{document}